\documentclass[journal]{IEEEtran}
\usepackage{graphicx} 
\usepackage[utf8]{inputenc}
\usepackage[T1]{fontenc}
\usepackage{cite}
\usepackage{amsmath}
\usepackage{algorithmic}
\usepackage[ruled]{algorithm}
\usepackage{array}
\usepackage[tight,footnotesize]{subfigure}
\usepackage{amsfonts}
\usepackage{multirow}
\usepackage{tikz}
\usepackage{stfloats}
\usepackage{url}
\usepackage{bm}
\usepackage{acronym}

\newcommand{\bfg}[1]{\boldsymbol{#1}} 
\newcommand{\bfb}[1]{\boldsymbol{\textrm #1}} 
\begin{document}

\title{Analytical Framework for Assessing Effective Regional Inertia}

\author
{
Bruno Pinheiro,
Joe H. Chow,
\IEEEmembership{Life Fellow,~IEEE,}
Federico Milano,
\IEEEmembership{Fellow,~IEEE,}
Daniel Dotta,
\IEEEmembership{Member,~IEEE}
\thanks{Bruno Pinheiro and Daniel Dotta are with the Department of Elec. Eng. at University of Campinas, Campinas, SP 13083-852, Brazil. e-mail: b229989@dac.unicamp.br; dottad@unicamp.br, 
Joe. H. Chow is with the Department of Elec. Comp. and Syst. Eng., Rensselaer Polytechnic Institute, Troy, NY 12180, USA. email: chowj@rpi.edu,
F. Milano is with the School of Elec. \& 
Electron. Eng., University College Dublin, Dublin, D04V1W8, Ireland. emails: federico.milano@ucd.ie. This work is supported by Sustainable Energy Authority of Ireland (SEAI) by funding F.~Milano under project FRESLIPS, Grant No.~RDD/00681, and by Coordination for the Improvement of Higher Education Personnel (CAPES) by funding Bruno P., Grant No.~001.}
}

\maketitle
\begin{abstract}

This paper proposes a novel formulation of effective regional inertia that explicitly accounts for both system topology and the spatial distribution of inertia. Unlike traditional approaches that model a region as an aggregated machine with an equivalent inertia, the proposed metric provides a topology-aware representation. The methodology builds on an analytical framework that extends classical slow coherency theory to address network partitioning and regional frequency stability. Based on these partitions, we develop a systematic procedure to evaluate the effective inertia of each region, enabling a more accurate interpretation of local inertial contributions, including those from virtual inertia provided by inverter-based resources (IBRs). Case studies on the IEEE 39-bus and 68-bus systems demonstrate that the integration of inertial devices does not uniformly improve system frequency response, underscoring the importance of the proposed metric for effective regional inertia assessment.

\end{abstract}


\section{Introduction}


The assessment of frequency stability has traditionally relied on conventional metrics such as the rate of change of frequency (RoCoF) and the frequency nadir, evaluated using the center-of-inertia (COI) frequency, or on aggregated models like system frequency response (SFR) \cite{Shi2018}. However, the dynamic behavior of frequency in power systems exhibits a spatiotemporal nature \cite{Milano_FDF}, a characteristic that becomes even more pronounced with the increasing heterogeneity introduced by IBRs. As consequence, conventional frequency stability assessment methods, such as those relying on the  COI frequency fail to capture localized transient behaviors effectively \cite{Zaichun2024,Bruno2025_SIN}. In this context, defining regions where frequency should be monitored and controlled, has become a critical aspect \cite{Badesa2021}. To properly address the challenges associated with evaluating regional frequency behavior, two key questions must be considered: (i) how to define \textit{coherent regions} for assessing regional frequency stability, and (ii) how to accurately quantify the \textit{effective regional inertia} while accounting for the spatial distribution of inertia within those regions. In this work we propose an unified analytical framework to define the regions and regional inertia assessment based on the concept of distribution of inertia. 


System partitioning groups generation and load buses into regions according to specific dynamic or operational objectives and is essential for stability assessment, control design, and resource planning in large-scale networks. The effectiveness of a partition depends critically on the chosen criterion (e.g., voltage regulation, static stability, or frequency dynamics)~\cite{Eddin2024}. For frequency studies, coherent-machine identification is classically addressed by slow-coherency theory~\cite{Chow_BookCoherency}, while extensions to include load buses rely on sensitivity factors~\cite{Chow_BookCoherency}. Measurement-based alternatives using measurements have also been proposed~\cite{DeCaro_EPSR_2023, Lugnani_TPWRS_2022}, but their dependence on event data and wide-area monitoring limits applicability in planning studies. 

In parallel, a number of studies have focused on regional inertia estimation using measurement-based techniques. In~\cite{Gotti2024}, generator internal frequencies are estimated and clustered, and an iterative system identification method is used to determine regional inertia. Reference~\cite{Tianshu2025} formulates a nonlinear parameter identification problem to estimate the equivalent inertia of a coherent area, using a limited number of frequency measurements and an inter-area frequency model. In~\cite{Lugnani_IJEPES2024}, transient frequency data is used to determine coherent regions and estimate regional inertia via a regional COI formulation. A similar effort in~\cite{Wilson2019} uses the average frequency of a predefined region and inter-area power exchange to estimate regional inertia, accounting for both generator and load contributions. Across these works, regional inertia is generally defined as the sum of inertia constants of machines and loads within a region, based on an equivalent swing equation representation. However, these approaches do not examine whether such equivalent metrics are physically meaningful for characterizing the actual regional inertial frequency response.

Beyond measurement-based estimation, recent works have emphasized the role of network connectivity and spatial inertia distribution in shaping regional frequency dynamics. To this end, screening approaches have been proposed that avoid full time-domain simulations and instead rely on analytical strength indices. In~\cite{Misyris2023}, regions are defined using synchronizing coefficients and generator inertia, with regional RoCoF metrics serving as stability indicators. In~\cite{Prasad2024}, inertia zones are formed using a maximal entropy random walk combined with a modified weighted k-means clustering. A frequency strength index is introduced in~\cite{Ignacio2025} to quantify the stability of coherent regions as a function of their inertial content. From a graph-theoretic perspective,\cite{Pagnier2019} reformulates generator frequency dynamics through the spectral decomposition of the system Laplacian matrix, showing that deviations are strongly tied to network topology, particularly the slowest eigenmodes. Similarly,\cite{Warren2022} employs Laplacian eigenvectors to identify coherent clusters and evaluates the regional COI frequency under different disturbance scenarios. However, methods relying solely on Laplacian properties inherently assume uniform inertia distribution and therefore require complementary metrics to capture spatial heterogeneity.

These studies highlight that regional vulnerability to frequency disturbances depends jointly on inertia distribution and network structure. As a result, there has been increasing interest in directly quantifying the spatial distribution of inertia across power systems~\cite{Painemal2018, fanhong2020final_letter, Prasad2023, Brahma2020}. For example,\cite{Painemal2018} introduces a time-domain-based inertia distribution metric, while\cite{Gayathri2023} proposes a variant based solely on PMU data. A simplified nodal inertia formulation is proposed in~\cite{fanhong2020final_letter} and later refined in~\cite{Bruno2023}, yielding the analytical formulation adopted in this work. Although these methods provide useful insights into locational or regional frequency stability, a clear gap remains: (i) there is still no unified framework that jointly integrates local frequency response, network topology, and coherent area definitions to analytically quantify regional inertial strength; and (ii) there is limited understanding of how the allocation of \textit{inertial devices}, that is, devices that provide inertial response such as synchronous generators (SGs), synchronous condensers (SCs), synchronous motors (SMs), or IBRs with virtual inertia, affects the regional inertial response.



The contributions of this paper are summarized as follows: 1) Building on the nodal inertia concept, we analyze the allocation of \textit{inertial devices} and demonstrate that their placement and contribution can, in certain cases, degrade the regional inertial response. We derive an analytical expression for the ideal inertia contribution as a function of device location. To the best of our knowledge, this is the first analytical formulation showing that adding inertia does not necessarily improve the regional inertial response; 2) We propose a novel analytical method for identifying coherency regions. The method combines nodal inertia with the network Laplacian matrix, enabling the inclusion of load buses in coherency assessments. Its main advantage lies in providing a more intuitive and interpretable framework for network partitioning, while maintaining direct comparability with classical slow-coherency-based approaches; 3) Building on the inertia distribution and identified coherent regions, we propose a new metric to quantify the regional inertia, namely \textit{effective regional inertia}. This metric facilitates the evaluation of how new inertial device affect the dynamics of the regional frequency response.




The subsequent sections of this paper are structured as follows: Section II provides the theoretical background. Section III presents the problem formulation. Section IV introduces the proposed approach to determine the regions, and the proposed effective regional inertia metric. Section V presents the study cases. Lastly, Section VI presents our concluding remarks.

\vspace{-0.3cm}

\section{Background}


\subsection{Fundamental Coherency in Power Systems}
\label{sec:Coherency}

The classical slow-coherency model aims to capture the swing dynamics and coupling among generators associated with low-frequency oscillation modes. The identification of coherent machines is based on a simplified representation of the power system. In this framework, coherent groups are determined independently of detailed machine models: SGs are represented by the classical second-order swing equation, and the effects of controllers are neglected \cite{Chow_BookCoherency}. After linearization around the steady-state operating point and elimination of algebraic variables, the basic electromechanical dynamics of $n_g$ generators can be expressed as \cite{Chow_BookCoherency}:
\begin{equation}
\label{eq:second_order_model}
\bfg{\Delta\dot{\omega}} = \bfb{M}^{-1} \bfb{K_s} \bfg{\Delta\delta} - \bfb{M}^{-1}\bfb{D}\bfg{\Delta\omega},
\end{equation}

\noindent where $\bfg{\Delta{\dot\omega}}\in \mathbb{R}^{n_g}$ is the vector of rotor acceleration deviations, $\bfg{\Delta{\delta}}\in \mathbb{R}^{n_g}$ is the rotor angle deviation, $\bfb{M}\in \mathbb{R}^{n_g \times n_g}$ is the diagonal inertia matrix, $\bfb{D} \in \mathbb{R}^{n_g \times n_g}$ is the diagonal damping/droop matrix, and $\bfb{K_s}\in \mathbb{R}^{n_g \times n_g}$ contains the synchronizing power coefficients (SPCs), which reflect the electrical coupling between machines. Assuming negligible conductance and applying Kron reduction, the entries of $\bfb{K_s}$ are given by:
\begin{equation}
\label{eq:KronNet}
K_{s_{\imath,\jmath}} = 
    \begin{cases}
        E_\imath E_\jmath B_{\imath,\jmath} \cos(\delta_\imath - \delta_\jmath), & \text{if } \imath \neq \jmath \\
        -\sum_{k \neq \imath} E_\imath E_k B_{\imath,k} \cos(\delta_\imath - \delta_k), & \text{if } \imath = \jmath
    \end{cases}
\end{equation}

\noindent where $B_{\imath,\jmath}$ is the equivalent susceptance between buses $\imath$ and $\jmath$, and $E_\imath$, $\delta_\imath$ are the internal voltage and angle of generator $\imath$.


In systems with IBRs providing frequency control, e.g., grid-following (GFL) inverters with frequency droop or grid-forming (GFM) devices with power–frequency droop, the damping term in \eqref{eq:second_order_model} cannot be neglected. This is because the effective damping is set by the droop gain $m_p$, with $D_{\text{IBR}} = 1/m_p$ \cite{Eugenie2024}. In this case, the system eigenvalues $\lambda$ are determined by the quadratic eigenvalue problem (QEP): $\textrm{det}(\lambda^2I + \bfb{M}^{-1}\bfb{D}\lambda -\bfb{M}^{-1} \bfb{K_s})=0$, and coherency must be assessed using the eigenvectors taking into account both inertia and damping/droop contributions. Coherent groups of generators are then identified by clustering based on the eigenvectors associated with the lowest-frequency modes.

\vspace{-0.4cm}

\subsection{Spectral Analysis}

A power system can be represented as an undirected weighted graph $\mathcal{G} = (\mathcal{V},\mathcal{E})$, where nodes $\mathcal{V}$ correspond to buses (generators and loads), and edges $\mathcal{E}$ represent transmission lines~\cite{Sanchez2014}. Edge weights vary with application, including line admittance, SPCs~\cite{Vladimir2013}, or power flow magnitudes~\cite{Sanchez2014}.

In lossless networks with $n = n_g+n_b$ buses, with $n_b$ being the number of load buses, the active power injected at bus $\imath$ is given by:
\begin{equation}
\label{eq:poweri}
    P_\imath = \sum_{\jmath \in \mathcal{E}} V_\imath V_\jmath B_{\imath\jmath} \sin(\theta_\imath - \theta_\jmath),
\end{equation}

\noindent where $V_\imath$ and $\theta_\imath$ are the voltage magnitude and angle at bus $\imath$, and $B_{\imath\jmath}$ is the susceptance of the line connecting buses $\imath$ and $\jmath$. Linearizing around an operating point yields the DC power flow model:
\begin{equation}
\label{eq:dcpower}
\begin{bmatrix}
\bfg{\Delta P_G} \\
\bfg{\Delta P_B}
\end{bmatrix}
=
\bfb{L}
\begin{bmatrix}
\bfg{\Delta \theta_G} \\
\bfg{\Delta \theta_B}
\end{bmatrix}
\end{equation}

\noindent where $\bfg{\Delta \theta_G}\in \mathbb{R}^{n_g}$ is the vector of generator buses angle, $\bfg{\Delta \theta_B}\in \mathbb{R}^{n_b}$ is the vector of load buses angle, and $\bfb{L} \in \mathbb{R}^{n \times n}$ is the network Laplacian matrix, defined as:
\begin{equation}
\label{eq:Laplacian}
L_{\imath\jmath} =
\begin{cases}
- V_{\imath} V_{\jmath} B_{\imath\jmath} \cos(\theta_\imath - \theta_\jmath), & \imath \neq \jmath \\
\sum_{k \neq \imath} V_\imath V_k B_{\imath k} \cos(\theta_\imath - \theta_k), & \imath = \jmath
\end{cases}
\end{equation}

The Laplacian matrix is symmetric and positive semi-definite. Its smallest eigenvalue $\lambda_1 = 0$ corresponds to a uniform angle shift, while the second smallest eigenvalue $\lambda_2$ is the \textit{algebraic connectivity}. The $k$ largest eigenvectors of $\bfb{L}$ form $\bfb{V} \in \mathbb{R}^{n \times k}$, with each row representing a node in $\mathbb{R}^k$. This mapping is called \textit{spectral embedding} \cite{Tyuryukanov2021}.

\vspace{-0.3cm}

\section{Problem Statement}

This section formalizes the problem addressed in this work, namely the regional inertia quantification. We begin with the formulation of the nodal inertia, followed by analyzing the trade-offs introduced by the integration of new inertial devices at the nodal level and then extend the discussion to the regional scale.

\vspace{-0.4cm}

\subsection{Nodal Inertia Formulation}

In this work, we consider the nodal inertia as the local frequency resistance following a power mismatch. The RoCoF at a particular bus $\jmath$ after a local disturbance $\Delta P_\jmath$ is thus defined as follows:
\begin{equation}
    \label{eq:rocof_busj}
    RoCoF_\jmath(t=0^+) = \frac{\Delta P_\jmath}{h_\jmath},
\end{equation}

\noindent where $h_\jmath$ represents the nodal inertia at bus $\jmath$. This underlying formulation highlights the role of nodal inertia in shaping local frequency variations, resulting in the following definition.

\vspace{0.15cm}

\noindent \textbf{Definition 1.} \textit{Nodal Inertia}: The nodal inertia of a bus $\jmath$, which may represent either a generation or load bus, is defined as the total inertial response observed in its local frequency deviation following a local power perturbation. Mathematically, nodal inertia is a weighted sum of the inertial contributions from all devices in the system that are dynamically linked to bus $\jmath$.

The quantification of nodal inertia begins with the formulation of bus frequencies. The \textit{frequency divider} (FD) expression provides an analytical estimate of the frequency deviations at each bus and is given by \cite{Milano_FDF}:
\begin{equation}
\label{eq:frequency_divider}
\bfg{\Delta{\omega}}_{b}(t) = \bfb{D_{\textrm{div}}} \bfg{\Delta{\omega}}(t),
\end{equation}

\noindent where $\bfg{\Delta{\omega}}_{b}(t)$ denotes the bus frequency deviations, and $\bfb{D_{\textrm{D}}} \in \mathbb{R}^{n \times n_g}$ is the \textit{frequency divider matrix}, defined as:
\begin{equation}
\label{eq:D_matrix}
\bfb{D_{\textrm{div}}} = \bfb{B_{BG}}^{+}[\bfb{B_{B}} + \bfb{B_{G}}],
\end{equation}

\noindent where $\bfb{B}_{B} \in \mathbb{R}^{n \times n}$ is the conventional susceptance matrix of the system; $\bfb{B}_{BG} \in \mathbb{R}^{n_g \times n}$ and $\bfb{B_G} \in \mathbb{R}^{n_g \times n_g}$ are the susceptance matrices associated with the generator internal nodes.

Following a power disturbance $\Delta P_\jmath$ at bus $j$, all synchronized generators instantaneously contribute active power via the energy stored in their magnetic fields. This power redistribution can be analytically described using the SPCs. The power contribution of generator $i$ is given by $\Delta P_i = \Delta S_{i,j} \Delta P_\jmath$, where $\Delta S_{i,j}$ denotes the linearized SPC of generator $i$ for a disturbance at bus $j$, expressed as:
\begin{equation}
    \label{eq:SPC}
    \Delta S_{\imath,\jmath} = \frac{\tilde{B}_{\imath,\jmath}E_{\imath_0}V_{\jmath_0}\cos(\delta_{\imath_0} - \theta_{\jmath_0})}{\sum\limits_{\imath = 1}^{n_g} \tilde{B}_{\imath,\jmath} E_{\imath_0}V_{\jmath_0}\cos (\delta_{\imath_0} - \theta_{\imath_0})}.
\end{equation}

\noindent Here, $E_{\imath_0}$ and $\delta_{\imath_0}$ denote the internal voltage and rotor angle of generator $\imath$ at the pre-disturbance operating point, while $V_{\jmath_0}$ and $\theta_{\jmath_0}$ correspond to the voltage magnitude and angle at bus $\jmath$.





The equivalent susceptance $\tilde{B}_{i,j}$ is obtained from an extended susceptance matrix that includes all generator buses and the selected bus $\jmath$. By applying Kron reduction to eliminate the remaining load buses (denoted by the subscript $r$), a reduced matrix is obtained, which preserves the coupling between the generators and bus $\jmath$. The generator–bus susceptance coupling vector can be calculated as
\begin{equation}
\label{eq:eqB}
\tilde{\bfg{B}}_{Gj} = B_{Gj} - B_{Gr} B_{rr}^{-1} B_{rj},
\end{equation}

\noindent where $B_{Gr}$ and $B_{rj}$ represent the generator–to–remaining-bus and remaining-bus–to–bus-$\jmath$ couplings, respectively, and $B_{rr}$ is the susceptance matrix of the eliminated buses. Repeating this procedure for each bus yields the complete matrix of equivalent susceptances,
\begin{equation}
\label{eq:concat_B}
\tilde{\bfb{B}} = \big[\tilde{\bfg{B}}_{G1}, \tilde{\bfg{B}}_{G2}, \ldots, \tilde{\bfg{B}}_{Gn}\big] \in \mathbb{R}^{n_g \times n_b}.
\end{equation}

Next, by taking the time derivative of \eqref{eq:frequency_divider}, substituting the swing equation~\eqref{eq:second_order_model}, the SPC definition in \eqref{eq:SPC}, and the underlying nodal inertia expression~\eqref{eq:rocof_busj}, one obtains the nodal inertia at bus $\jmath$ \cite{Bruno2023}:
\begin{equation}
    \label{eq:hj}
    h_\jmath = \dfrac{1}{\sum_{k=1}^{n_g} \frac{1}{2H_k} D_{\textrm{div}_{\jmath,k}} \Delta S_{k,\jmath}} .
\end{equation}

It should be noted that the damping/droop terms of the devices do not affect the nodal inertia. To generalize \eqref{eq:hj} for all buses, the following matrices are defined:
\begin{align}
    \bfb{K}   &= \bfg{E_g}\bfg{1}_{n}^{\top} \circ \left( \tilde{\bfb{B}} \circ \cos(\bfg{\Delta}) \right), \label{eq:K} \\
    \bfg{K_h} &= \bfg{1}_{n_g}^\top \bfb{K}, \label{eq:Kh}
\end{align}

\noindent where $\circ$ denotes element-wise multiplication; $\bfb{K}\in \mathbb{R}^{n_g \times n}$; $\bfg{E_g}$ is the vector of generator internal voltages; $\bfg{\Delta} \in \mathbb{R}^{n_g \times n}$ is the matrix of angle differences between $\delta$ and $\theta$; and $\bfg{1}_{n_g}$ is a unit vector of dimension $n_g$.

The combined effect of the FD matrix and generator inertia is then captured by the frequency sensitivity matrix $\bfb{F}\in \mathbb{R}^{n \times n_g}$, defined as:
\begin{equation}
\label{eq:F}
    \bfb{F} = \left(\bfb{D_{\textrm{div}}} \circ {\bfb{K}}^\top  \right) \bfb{M}^{-1},
\end{equation}

\noindent  where $\bfb{M}$ is the diagonal matrix with generators inertia.

Defining:
\begin{equation}
\label{eq:Fh}
    \bm{F_h} = \bfb{F} \bfg{1}_{n_g},
\end{equation}

\noindent and finally, combining \eqref{eq:Kh} and \eqref{eq:Fh}, the nodal inertia vector $\bfg{h} \in \mathbb{R}^{n \times 1}$ is obtained as:
\begin{equation}
\label{eq:final_h}
    \boxed{\bfg{h} = \bfg{K}_h^{\top} \oslash \bfg{F}_h}.
\end{equation}

\noindent where $\oslash$ denotes element-wise division. Equation~\eqref{eq:final_h} provides the closed-form analytical expression for nodal inertia at each bus. 

In summary, the nodal inertia distribution is governed by the generator parameters, the system topology, and the operating point. The inertial contribution of IBRs or loads with inertial response\footnote{This contribution is restricted to synchronous devices that provide an effective instantaneous inertial response, such as SMs.} can be direct included in the formulation \eqref{eq:final_h}. The impact of virtual inertia from GFM devices are presented later in Section \ref{sec:case_study_1}.

\subsubsection{Distribution of Damping}

As a byproduct of the proposed framework, the distribution of damping can also be determined. For $t > 0^+$ and prior to the activation of SGs frequency controls, the damping distribution is obtained analogously to \eqref{eq:F}, and is defined as
\begin{equation}
    \bfb{R_{\textrm{d}}} = \left(\bfb{D_{\textrm{div}}} {\bfb{D}}  \right) \bfb{M}^{-1}\in \mathbb{R}^{n\times n_g}, 
\end{equation}

\noindent leading to
\begin{equation}
\label{eq:R_droop}
      \bfb{R} = \textrm{diag}( \bfb{R_{\textrm{d}}}\bfg{1}_{n_g}) \in \mathbb{R}^{n\times n}.
\end{equation}

A detailed examination of damping distribution lies beyond the scope of this work. Instead, the analysis here focuses on regional inertia assessment, with the formulation in \eqref{eq:R_droop} applied specifically to coherency studies in the presence of IBRs operating under droop-based frequency control.

\vspace{-0.3cm}

\subsection{Inertia Allocation Trade-Offs}
\label{sec:example}

\begin{figure}[ht]
\includegraphics[width=0.85\linewidth]{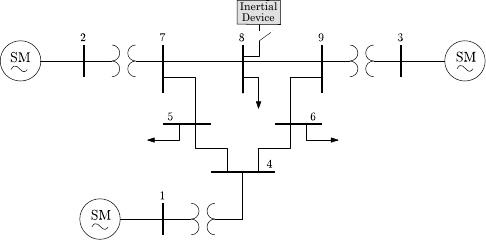}
\centering
\caption{WSCC 9-bus system with a new inertial device connected at bus~8.}
\label{fig:9bus_diagram}
\end{figure}

The integration of an inertial device directly modifies the local and system-wide inertia response. To illustrate this, we consider the WSCC 9-bus system, where an inertial device with inertia constant $H_\mathrm{device}$ is connected at bus~8, as shown in Fig.~\ref{fig:9bus_diagram}.  

\begin{figure}[htb]
\includegraphics[width=1\linewidth]{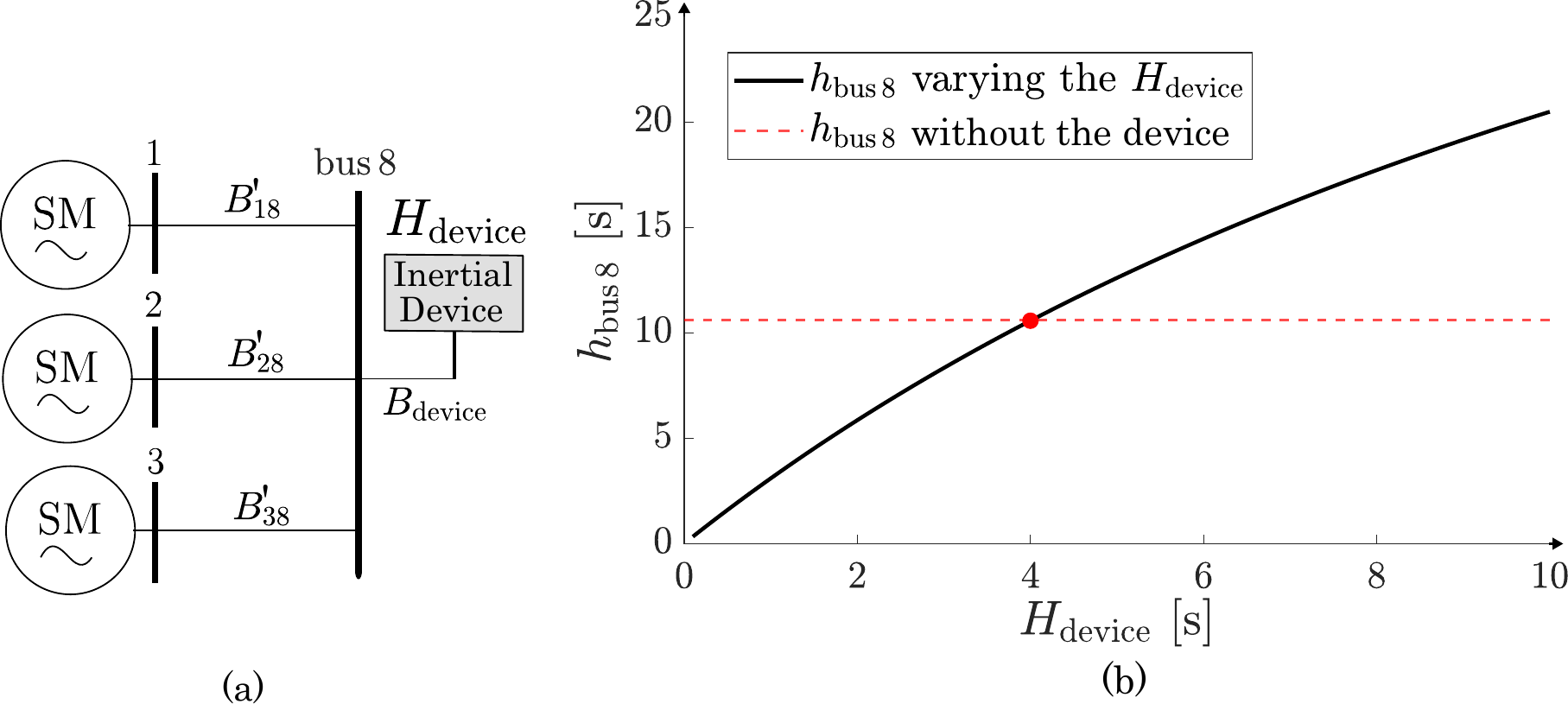}
\centering
\caption{(a) Equivalent system seen from bus 9, and (b) the nodal inertia at bus~8 as a function of the device inertia constant $H_\mathrm{device}$.}
\label{fig:varyingh8}
\end{figure}

The nodal inertia at bus~8 is computed using~\eqref{eq:final_h}. Fig.~\ref{fig:varyingh8}a depicts the equivalent representation seen from bus~8, while Fig.~\ref{fig:varyingh8}b shows the nodal inertia trajectory as $H_\mathrm{device}$ varies from $0.1\,\mathrm{s}$ to $10\,\mathrm{s}$. The dashed red line corresponds to the base case without the additional device. Results show that the post-connection nodal inertia can be either lower or higher than the base case, depending on $H_\mathrm{device}$. Improvement is observed when $H_\mathrm{device} > 4\ \mathrm{s}$, as indicated by the red dot.  

The observed behavior arises because the new device connection redistributes the inertial contributions of the existing SG. Due to the strong electrical proximity of the new device, i.e., $B'_{\mathrm{device}} \ll B'_{1,8},B'_{2,8},B'_{3,8}$ and $D'_{8,4}\gg D'_{8,1},D'_{8,2},D'_{8,3}$, the effective contribution of the SGs to bus~8 decreases, and the new device dominates the local inertia response.  

\begin{figure}[htb]
\includegraphics[width=0.8\linewidth]{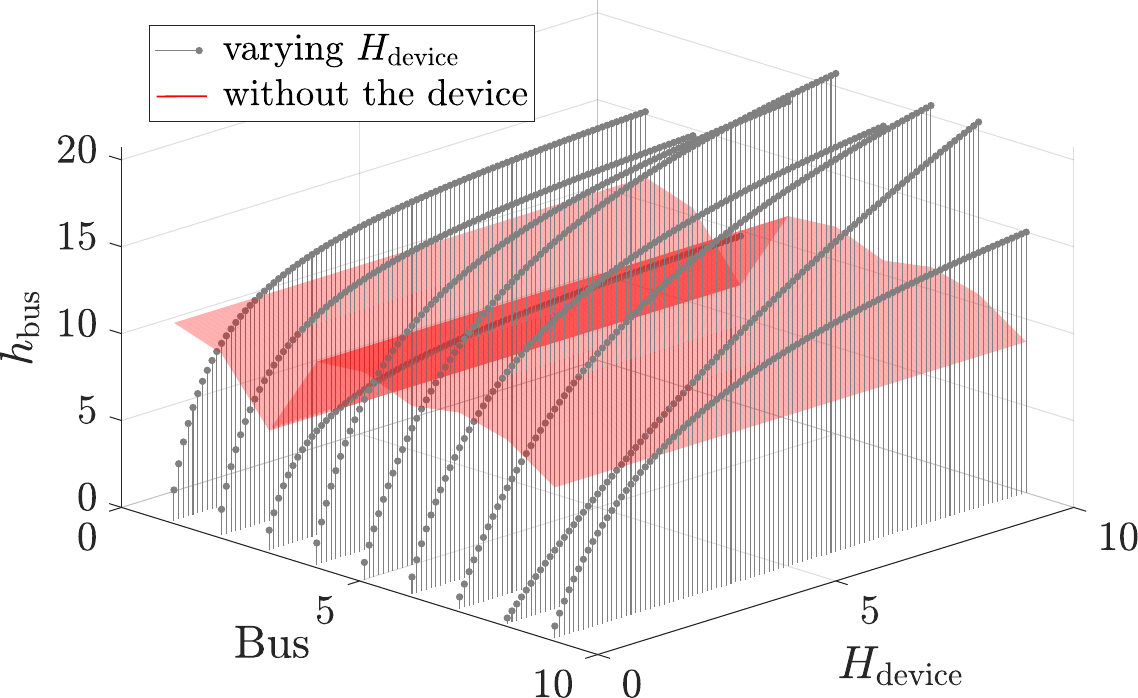}
\centering
\caption{Nodal inertia distribution across all buses as a function of $H_\mathrm{device}$.}
\label{fig:allbuses_9bus}
\end{figure}

The impact of this redistribution is not restricted to the connection bus. Fig.~\ref{fig:allbuses_9bus} shows the system-wide nodal inertia profile as $H_\mathrm{device}$ varies. In this reduced system, certain $H_\mathrm{device}$ values lead to a reduction of nodal inertia across several buses, highlighting the system-wide implications of local device placement and its inertial capacity.  

Additionally, Fig. \ref{fig:allbuses_9bus} shows the impact of the variation of the device's inertia on the distribution of inertia across the system. We can note that for some $H_{\mathrm{device}}$ values, the distribution of inertia across all the system is reduced. This is specially critical for this reduced system, where the connection of the inertial device impact the hole system.

Based on this discussion, the following remark is relevant:

\vspace{0.15cm}
\noindent \textbf{Remark 1}: The connection of an inertial device does not guarantee an enhancement of the system inertial response. Its impact depends both on the inertia constant of the device and on the pre-existing inertia distribution of the network.  

\subsubsection{Illustrative Example}

\begin{figure}[ht]
\includegraphics[width=0.85\linewidth]{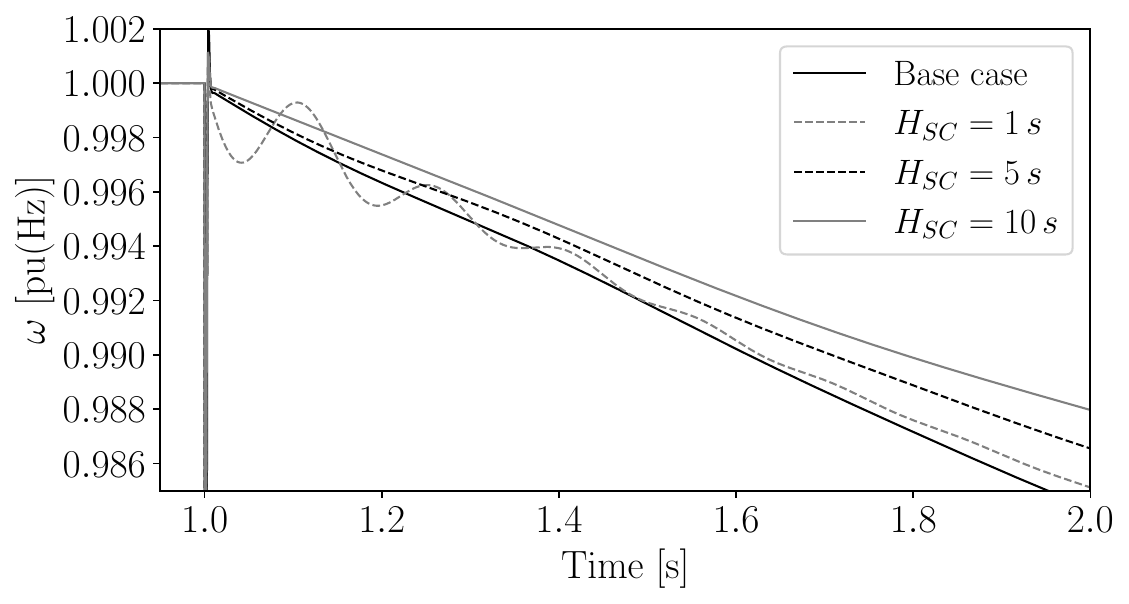}
\centering
\caption{Frequency response at bus 8 under a 1 p.u. load step for different inertia scenarios connected at the same bus.}
\label{fig:freq_9bus}
\end{figure}

As an illustrative case, let us consider the inertial device as a SC. Three cases of inertia constant $H_{\mathrm{SC}}$ are analyzed: $1\,\mathrm{s}$, $5\,\mathrm{s}$, and $10\,\mathrm{s}$. To evaluate the inertial response, we apply a 1 p.u load step at bus 4 for each case. The local frequency at bus 8 is measured using a conventional synchronous reference frame Phase Locked Loop (PLL). The frequency response for each scenario is shown in Fig.~\ref{fig:freq_9bus}. Among the scenarios, the case with $H_{\mathrm{SC}}=1\,\mathrm{s}$ exhibits the highest local RoCoF, that is, its inertial response is worse than the base case without this new device, corroborating Remark 1. 

The reduction in nodal inertia can be attributed to the introduction of a new local oscillation mode when the device operates with low inertia. This mode excites fast dynamics in the system. Although the analytical formulation of nodal inertia \textbf{is not modal-based}, the impact of the local mode is inherently reflected in the lower inertia values in the connection bus. The local inertial response is determined not only by the inertia constants of individual devices but also by their dynamic coupling within the grid

\vspace{-0.3cm}

\subsection{Regional Inertia Quantification}

\begin{figure}[t]
\includegraphics[width=0.65\linewidth]{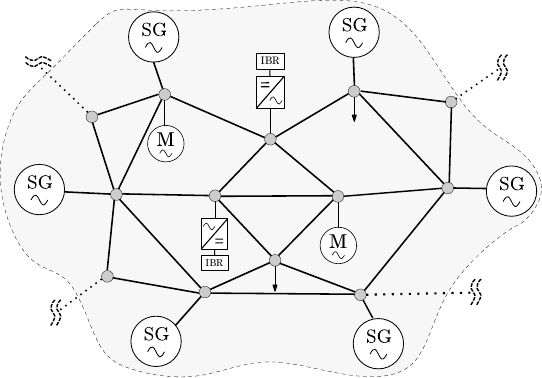}
\centering
\caption{Power system region with the location of a new device highlighted.}
\label{fig:region}
\end{figure}

While the nodal perspective is useful for understanding local allocation, inertia must also be assessed at a regional level in large-scale systems. Consider a generic region $\mathcal{R}$, as shown in Fig.~\ref{fig:region}. A conventional approach to quantify its inertia is given by  
\begin{equation}
    \label{eq:conventional_H}
    \mathrm{H}_{\text{conv}}^{\mathcal{R}} = \sum_{\jmath\in\mathcal{R}_{\mathrm{sm}}}H_\jmath + \sum_{\imath\in\mathcal{R}_{\mathrm{ibr}}}H_\imath + \sum_{d\in\mathcal{R}_{\mathrm{dev}}}H_d,
\end{equation}

\noindent where $\mathcal{R}_{\mathrm{sm}}$ denotes the set of synchronous machines, $\mathcal{R}_{\mathrm{ibr}}$ the set of IBRs providing virtual inertia, and $\mathcal{R}_{\mathrm{dev}}$ the set of other inertia-contributing devices within region $\mathcal{R}$ (e.g., synchronous motors or other dynamic loads).

Although restricted to a region, this formulation effectively reduces the entire subsystem to a lumped equivalent, ignoring both internal topology and the spatial distribution of inertia sources. Consequently, it assumes that any new device connected within the region contributes uniformly to the regional inertia. As highlighted in the previous subsection, this assumption is often misleading, since the location and the inertial capacity of the device within the network critically shapes the regional frequency behavior.  

To address these limitations, the problem considered in this work is formulated in two stages: (i) Partition the system into coherent regions based on frequency dynamics and nodal inertia distribution, ensuring that both generator and load buses are meaningfully represented; (ii) Define a new \textit{effective regional inertia} metric that accounts not only for the total inertial content of the devices in the region but also for its spatial allocation and interaction with network topology.  

The first stage leverages principles from classical slow coherency theory but extends them to include load buses, which are traditionally neglected in coherency analysis. The second stage develops a representative inertia measure for each region that reflects both aggregate inertia and its distribution. This two-stage framework provides the foundation for a more accurate assessment of inertia adequacy in a regional level.  

\vspace{-0.3cm}

\section{Proposed Framework}

This section introduces the proposed framework for effective regional inertia assessment. First, the network partitioning approach is described. Then, the procedure for quantifying regional inertia is presented, followed by a discussion on determining the minimum inertia required from a newly allocated device to enhance the effective regional inertia.

\vspace{-0.4cm}

\subsection{Network Partitioning}
\label{sec:netpart}

The classical slow coherency model presented in Section~\ref{sec:Coherency} relies on a reduced network representation in which only generator internal nodes are retained. In this framework, the system is modeled as a weighted graph $\mathcal{G} = (\mathcal{V}, \mathcal{E}, \mathcal{W})$, where $\mathcal{V}$ denotes the set of generator nodes, $\mathcal{E}$ represents the electrical couplings among them, and $\mathcal{W}$ encodes nodal weights associated with generator inertia and damping/droop coefficients. 

For regional frequency assessment, however, it is essential to preserve the full network structure so that coherent regions of buses can be identified, i.e., through \emph{network partitioning}. To overcome this limitation, we extend the reduced generator graph to a full-bus graph, denoted by $\mathcal{G} = (\mathcal{B}, \mathcal{L}, \mathcal{N})$, where $\mathcal{B}$ is the set of buses, $\mathcal{L}$ captures the electrical couplings, and $\mathcal{N}$ defines nodal weights that reflect the distribution of both inertia and damping/droop across the network. This extended representation enables the direct application of spectral methods for system partitioning at the load bus level.  

For a given operating point, the electrical coupling of system buses is represented by the Laplacian matrix $\bfb{L}$, as defined in~\eqref{eq:Laplacian}. The nodal inertia vector $\bfg{h}$ is obtained from~\eqref{eq:final_h}, and the corresponding diagonal inertia matrix is defined as $\bfb{N} = \mathrm{diag}(\bfg{h})$. Likewise, the damping contributions from synchronous generators and the power--frequency (P--f) droop control of IBRs are aggregated into the diagonal matrix $\bfb{R}$, as defined in~\eqref{eq:R_droop}.  

The spectral analysis of the extended model leads to the QEP:  
\begin{equation}
    \label{eq:QEP}
    \big(\bar{\lambda}^2 \bfb{N} + \bar{\lambda} \bfb{R} + \bfb{L}\big)\bar{\phi} = \bfb{0},
\end{equation}
where $\bar{\lambda}\in \mathbb{C}$ denotes an eigenvalue and $\bar{\phi}\in \mathbb{C}^{n_b}$ the corresponding eigenvector. Since $\bfb{N} \succ 0$ (positive definite), the QEP is well-posed and admits $2n_b$ eigenvalues. 
 
Once the eigenvalues and eigenvectors of~\eqref{eq:QEP} are obtained, a spectral clustering procedure is applied to determine the network partitioning. Specifically, the eigenvalues are organized in ascending order, i.e., $|\bar{\lambda}_1|<|\bar{\lambda}_2|\dots<|\bar{\lambda}_{n_b}|$, and the spectral embedding is formed by stacking the first $k$ associated eigenvectors into a matrix $\bfb{V}_s \in \mathbb{C}^{n_b \times k}$, which maps each bus from the $n_b$-dimensional network space to a $k$-dimensional spectral space. The underlying regions of the system are then revealed by applying a clustering algorithm to the rows of $\bfb{V}_s$. 

The number of eigenvectors $k$ to be used in the spectral embedding is determined using the \textit{eigengap heuristic}, which identifies the largest relative gap between consecutive eigenvalues \cite{Sanchez2014}:
\begin{equation}
\label{eq:eigengap}
    \gamma_i = \frac{|\bar{\lambda}_{i+1}| - |\bar{\lambda}_i|}{|\bar{\lambda}_i|}.
\end{equation}

The index $k$ corresponding to the maximum $\gamma_i$ indicates the optimal number of significant eigenvectors to represent the coherent structure. This heuristic is based on the principle that well-separated clusters are associated with small eigenvalues, while larger eigenvalues correspond to weaker, less coherent modes. It thus identifies the eigenvectors that capture the dominant coherent dynamics, discarding those associated with minor variations.

Finally, the k-means clustering algorithm is applied in $\bfb{V}_s$ to identify the coherent regions. The choice of k-means is motivated by its simplicity and effectiveness when the data exhibits clearly separable clusters due to the incorporation of both electrical distances and the nodal inertia distribution. The number of clusters $r$ must be specified \textit{a priori} and can be determined using well-defined metrics, e.g., the silhouette score, which evaluates the compactness and separation of the resulting clusters.

When damping is neglected ($\bfb{R} = \bfb{0}$),~\eqref{eq:QEP} reduces to $(\bar{\lambda}^2 \bfb{N} + \bfb{L})\bar{\phi} = \bfb{0}$. Introducing $\mu = \bar{\lambda}^2$, this is equivalent to the generalized eigenvalue problem $\bfb{L}\bar{\phi} = -\mu \bfb{N}\bar{\phi}$, where the spectrum $\{\mu_i\}$ is real and nonnegative. Hence, the eigenvalues of the QEP appear in conjugate purely imaginary pairs, $\bar{\lambda} = \pm j\sqrt{\mu_i}$, with $0=\mu_1  < \mu_2 \leq \dots \leq \mu_{n_b}$. In this case, the spectral properties are fully characterized by the matrix pencil $(\bfb{L}, \bfb{N})$.

The extended spectral formulation presented in this section preserves the interpretation of slow coherency while enabling clustering of \emph{all} buses, thereby supporting regional frequency assessment in networks with heterogeneous inertia and damping distributions, including systems with high IBR penetration.



\vspace{-0.5cm}

\subsection{Effective Regional Inertia}
\label{sec:freqeval}

In the following, we propose a novel definition to quantify the regional inertia related to a coherent-regions determined by the extended slow coherency. The effective regional inertia metric can be calculated reflecting the distribution of inertia within the region and the system topology, as follows:

\vspace{0.15cm}

\noindent \textbf{Definition 2.} \textit{Effective Regional Inertia}: Let $\mathcal{R}$ denote a region identified through the extended slow coherency framework. The effective regional inertia is defined as the average nodal inertia within that region:
\begin{equation}
\label{eq:RI}
\mathrm{H}_{\text{eff}}^\mathcal{R} = \frac{\sum_{i \in \mathcal{R}} h_i}{n_\mathcal{R}},
\end{equation}

\noindent where $h_i$ is the nodal inertia at bus $i$, and $n_\mathcal{R}$ is the total number of buses in region $\mathcal{R}$.




\vspace{-0.5cm}

\subsection{Minimum Inertia of a Device}
\label{sec:minH}

Building upon the previous discussion, we now derive a criterion to determine the minimum inertia required from a new inertial device such that its connection does not degrade the local inertial response at its point of connection. 

Let us consider the nodal inertia at a generic bus $\jmath$ before the installation of the new device, expressed as:
\begin{equation}
    \label{eq:minH_old}
    \frac{1}{h_\jmath^{\mathrm{old}}} = \sum_{k=1}^{n_g} \frac{1}{2H_k} D_{j,k} \Delta S_{k,j}.
\end{equation}
After connecting a new device with (virtual) inertia $H_{\mathrm{device}}^j$ directly at bus $\jmath$, the updated nodal inertia becomes:
\begin{equation}
\label{eq:minH_new}
    \frac{1}{h_\jmath^{\mathrm{new}}} = \frac{1}{2H_{\mathrm{device}}^j} D_{j,\mathrm{device}} \Delta S_{\mathrm{device},j} + \sum_{k=1}^{n_g} \frac{1}{2H_k} D'_{j,k} \Delta S'_{k,j}
\end{equation}
where $\bfb{D}'_{j,k}$ and $\Delta S'_{k,j}$ denote the updated FD matrix and SPC terms after the addition of the new inertial device. It is important to note that the FD matrix is row-normalized; therefore, the inclusion of a new, electrically close device reduces the relative weightings of distant generators in the updated sum.

To ensure that the local inertial response is preserved or improved, the following condition must hold:
\begin{equation}
\label{eq:condition}
    h_\jmath^{\mathrm{new}} \geq h_\jmath^{\mathrm{old}} \Leftrightarrow \frac{1}{h_\jmath^{\mathrm{new}}} \leq \frac{1}{h_\jmath^{\mathrm{old}}}
\end{equation}

Define the shorthand terms: $\mathcal{F}_{k,\jmath} = D_{\jmath,k} \Delta S_{k,\jmath}$, $\mathcal{F}'_{k,\jmath} =  D'_{\jmath,k} \Delta S'_{k,\jmath}$, and $\mathcal{F}_{\jmath,\mathrm{device}} =  D_{\jmath,\mathrm{device}} \Delta S_{\mathrm{device},\jmath}$. Substituting \eqref{eq:minH_old} and \eqref{eq:minH_new} into \eqref{eq:condition}, the minimum (virtual) inertia required from the device at bus $\jmath$ is given by:
\begin{equation}
\label{eq:min_Hj_criterion}
\boxed{H_{\mathrm{device}}^\jmath \geq  \frac{ \mathcal{F}_{\jmath,\mathrm{device}} }{\sum\limits_{k=1}^{n_g} \left( \frac{\mathcal{F}_{k,\jmath} - \mathcal{F}'_{k,\jmath}}{H_k} \right)}}
\end{equation}

Equation~\eqref{eq:min_Hj_criterion} provides a analytical and straightforward constraint for designing virtual or physical inertia such that the local inertial response at the point of connection is preserved or enhanced. It highlights that the added inertia must compensate for the redistribution of FD weight and SPC contributions, which naturally occurs when new devices are introduced into the system topology. 

Based on the minimum inertia formulation in~\eqref{eq:min_Hj_criterion} and the discussion of this previous section, the following remark is relevant:

\noindent \textbf{Remark 2}: The contribution of inertia from a newly allocated device within a well-defined 
region depends on the pre-existing inertia level of that region. For a device connected at bus $j$, assuming $\mathcal{F}_{\jmath,\mathrm{device}} \approx 1$ and $\mathcal{F}'_{k,\jmath} \approx 0$, the condition $H_{\textrm{device}} \geq h_j^{\textrm{old}}$ holds, i.e., the device inertia must be greater than the inertia of the connection bus prior to allocation. This condition corresponds to an edge case and should be interpreted as a conservative bound for analysis.


The formulation in \eqref{eq:min_Hj_criterion} applies when a new device is added without disconnecting existing machines. Cases where nodal inertia at the connection bus increases due to other factors are analyzed in Section~\ref{sec:casestudy}.

\vspace{-0.4cm}

\section{Case Study}
\label{sec:casestudy}

This section discusses the proposed quantification of effective regional inertia in different scenarios using  IEEE 39-bus and 68-bus benchmark systems. The time domain simulations results are obtained using the power system analysis software tool Dome \cite{Dome}.

\vspace{-0.3cm}

\subsection{IEEE 39-bus Benchmark}
\label{sec:case_study_1}

The first case study considers the standard IEEE 39-bus benchmark system, which consists of ten synchronous machines. In the simulations, all machines are modeled using a fourth-order representation. Detailed parameters for the generators and their associated controllers, as well as the system’s initial operating point, are provided in~\cite{Benchmark}. The nodal inertia values discussed here are referenced to the system base of 100~MW.

\subsubsection{Distribution of Nodal Inertia and Regions}

\begin{figure}[htb]
\includegraphics[width=0.70\linewidth]{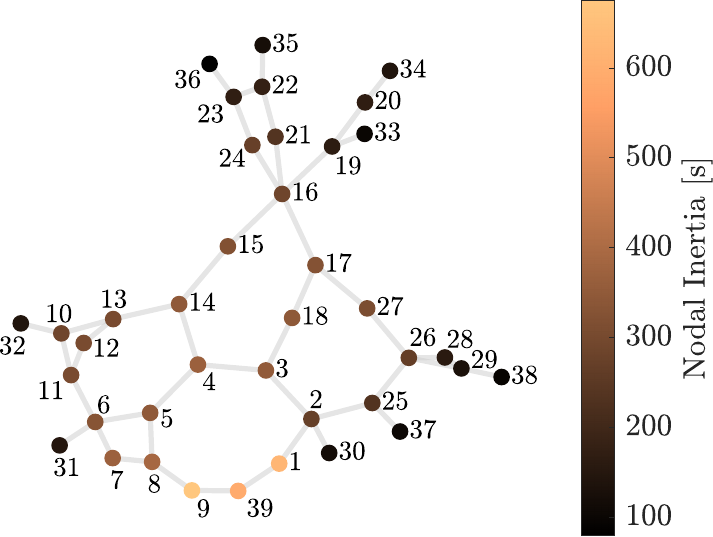}
\centering
\caption{Nodal inertia distribution in the IEEE 39-bus system for the base case.}
\label{fig:nodal_39bbench}
\end{figure}

The first step is the calculation of the distribution of inertia, which is shown in Fig.~\ref{fig:nodal_39bbench}. The buses with the highest nodal inertia are buses 1, 9, and 39, with bus 9 exhibiting the highest nodal inertia -- corresponding to an inertial response of approximately $700\,\mathrm{s}$. In contrast, the generator buses with the lowest nodal inertia are buses 33, 36, and 38, with bus 36 presenting an inertial response of approximately $80\,\mathrm{s}$.

\begin{figure}[ht]
\includegraphics[width=0.90\linewidth]{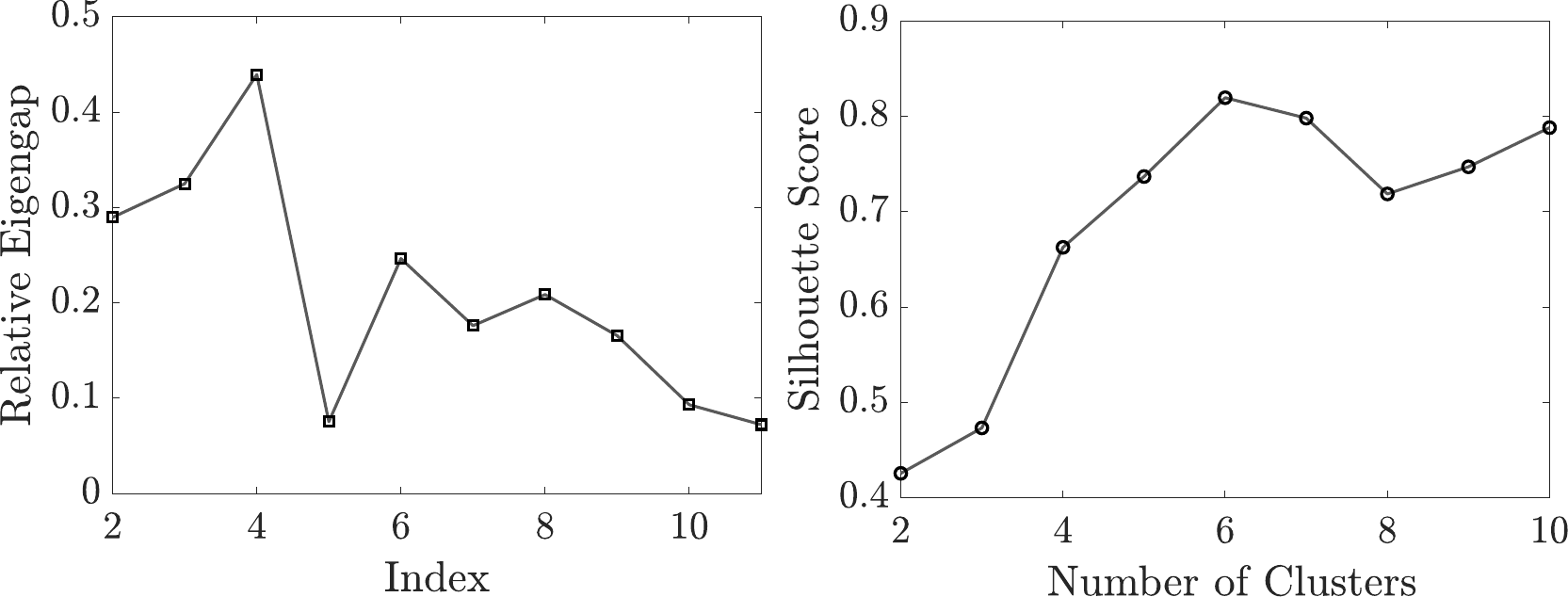}
\centering
\caption{Relative eigengap (left) and silhouette coefficient (right) for the IEEE 39-bus system.}
\label{fig:metrics39b}
\end{figure}

\begin{figure}[ht!]
\includegraphics[width=0.95\linewidth]{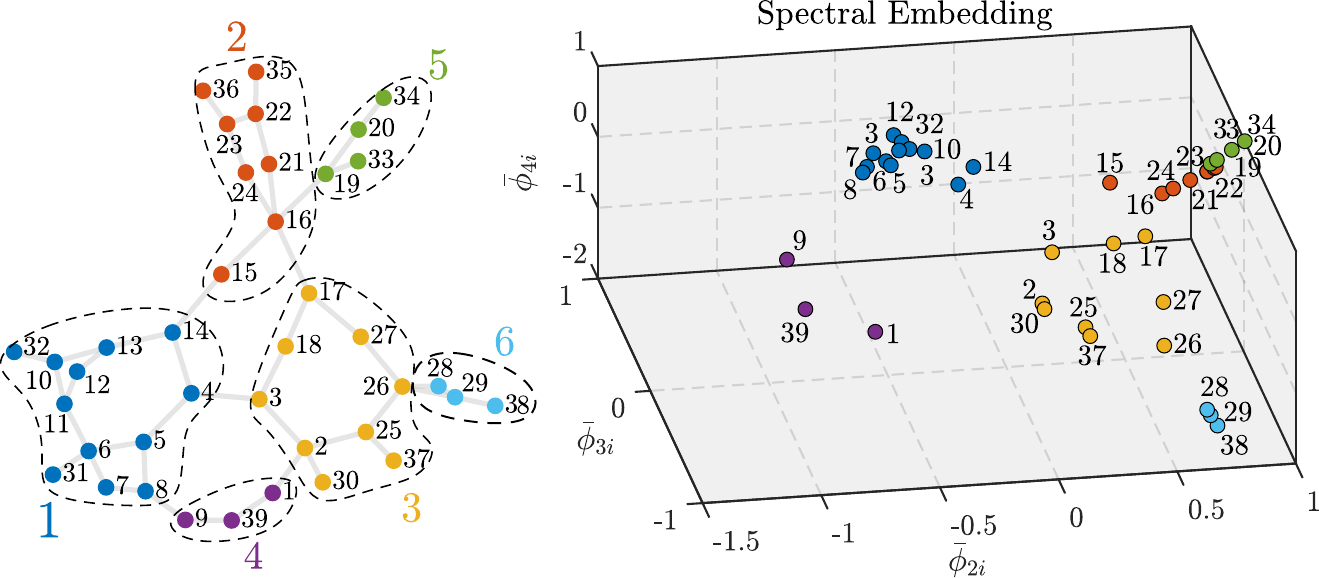}
\centering
\caption{Network partitioning and spectral embedding of the IEEE 39-bus system using the three dominant eigenvectors of the pencil  $(\bfb{L},\bfb{N})$.}
\label{fig:clusters_39bbench}
\end{figure}

The regions are determined by the network partitioning approach described in Sec. \ref{sec:netpart}. Fig.~\ref{fig:metrics39b} shows that the \textit{relative eigengap} metric indicates that four eigenvectors should be used to form the \textit{spectral embedding}. Accordingly, the four eigenvectors associated with the four smallest non-zero eigenvalues are used to determine the clusters. The Silhouette Coefficient is also show in Fig.~\ref{fig:metrics39b}, indicating that the optimal number of coherent regions is six. Finally, the k-means algorithm is applied to define the six coherent regions. The resulting regions are depicted in Fig.~\ref{fig:clusters_39bbench}. The obtained regions align with those reported in previous studies such as~\cite{Warren2022,Yusof1993}. 

To better visualize how the obtained eigenvectors from the pencil reflect the clustering structure, the \textit{spectral embedding} using three of these eigenvectors is shown in Fig.~\ref{fig:clusters_39bbench}, where node colors correspond to their assigned regions. It is worth to note how the combination of system topology information and the nodal inertia allows the clear separation between the regions.

\subsubsection{Effective Regional Inertia}

\begin{figure}[htb]
\includegraphics[width=0.85\linewidth]{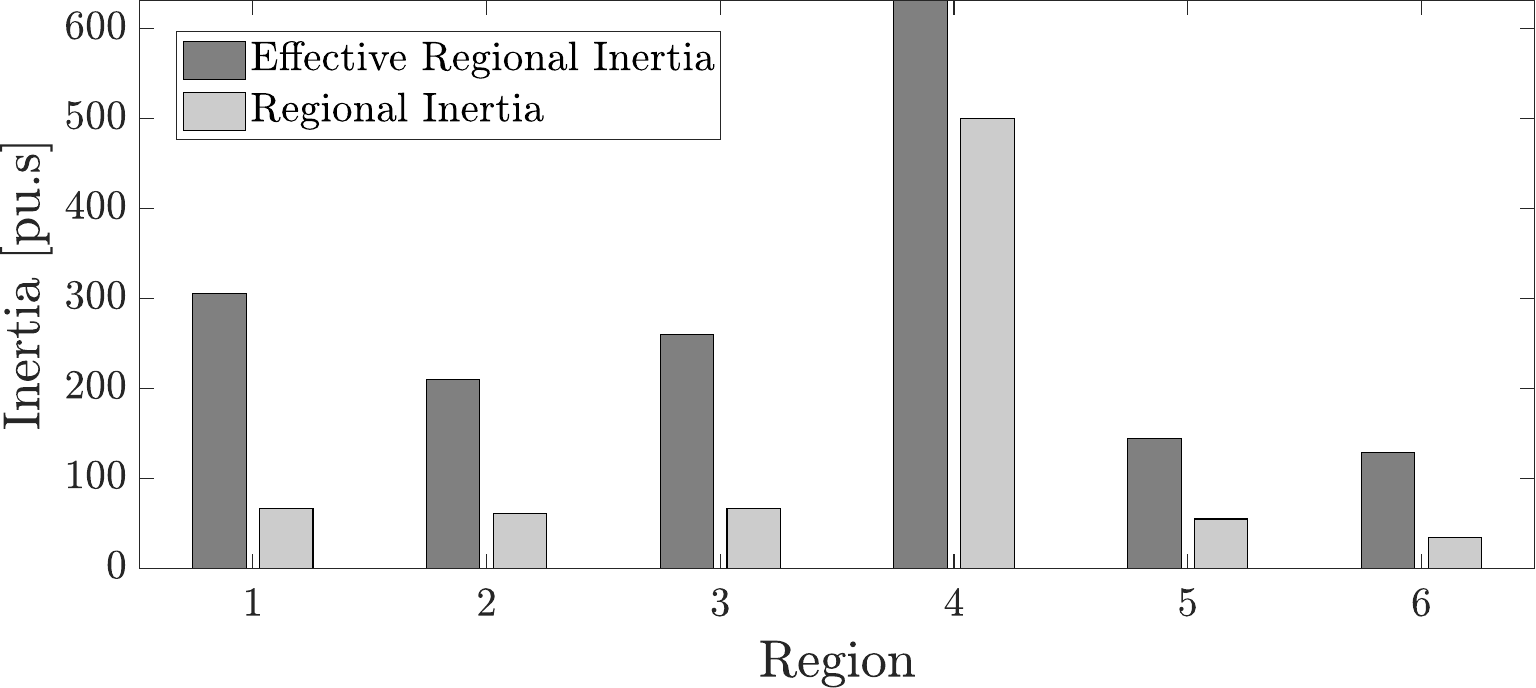}
\centering
\caption{Effective and conventional regional inertia in the IEEE 39-bus system.}
\label{fig:regionalinertia_39b}
\end{figure}

Based on the obtained regional partitions and the distribution of nodal inertia, the effective regional inertia is computed using~\eqref{eq:RI}, as illustrated in Fig.~\ref{fig:regionalinertia_39b}. For comparison purposes, the conventional regional inertia is also calculated using~\eqref{eq:conventional_H}. Among all regions, Region 4 exhibits the highest effective regional inertia, while Region 6 presents the lowest. In contrast, the conventional regional inertia values are relatively uniform across most regions, except for Region~4, which stands out with a significantly higher value.

\begin{figure}[htb]
\includegraphics[width=0.95\linewidth]{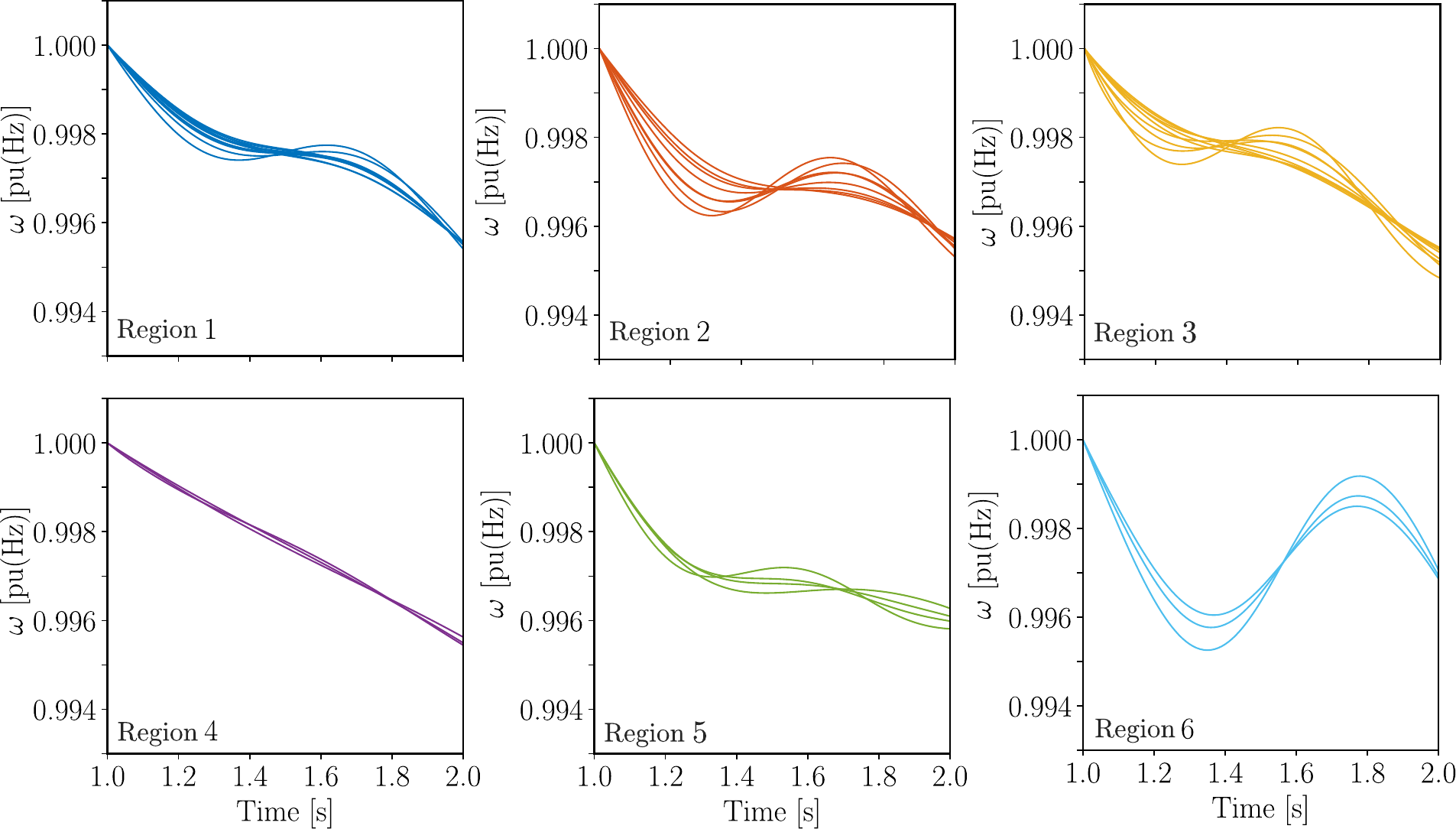}
\centering
\caption{Frequency dynamics of each region in the IEEE 39-bus system following a local load disturbance applied within that same region.}
\label{fig:freqs_39b_regions}
\end{figure}

To validate these results, a positive load step corresponding to approximately 15\% of the system's total demand is applied individually to each region. The disturbance is introduced at the bus with the lowest nodal inertia within each region (buses 10, 23, 25, 4, 1, 5, 19, 6, and 29). The corresponding frequency responses at all system buses are presented in Fig.~\ref{fig:freqs_39b_regions}.

The resulting frequency responses demonstrate clear internal coherency within each of the identified regions, thereby validating the effectiveness of the regional partitioning approach. Furthermore, the disturbance applied to Region 6 results in the largest frequency deviations and RoCoF values within that region, consistent with its low effective inertia. Conversely, Region 4 exhibits the smallest frequency deviations following the disturbance. These findings support the interpretation of effective regional inertia as a reliable indicator of locational frequency strength. Additionally, noticeable differences are observed in the frequency behavior among Regions 1, 2, and 3. In particular, Region 2 experiences the highest frequency deviations and initial RoCoF, whereas Region 1 ans 3 exhibits more moderate responses. However, these differences are not captured by the conventional regional inertia metric.

\subsubsection{Impact of inter-regional contributions}

\begin{figure}[htb]
\includegraphics[width=0.7\linewidth]{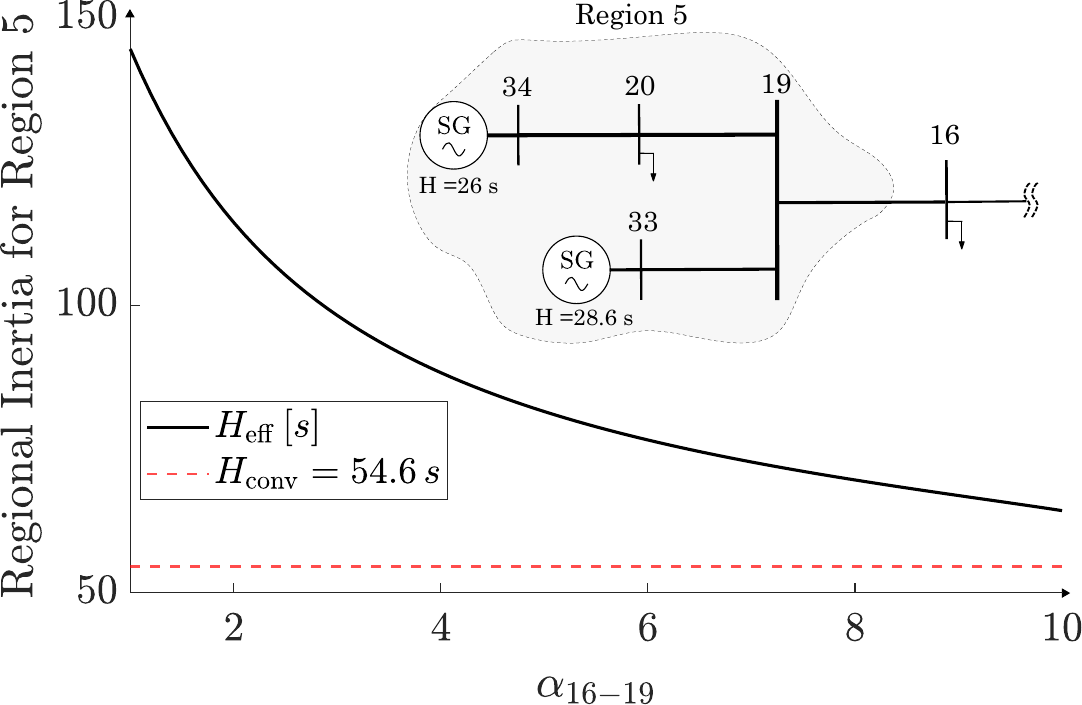}
\centering
\caption{Comparison between the proposed and the conventional regional inertia for Region 5 under variations of the interconnection line reactance.}
\label{fig:varyline_region5}
\end{figure}

The results in Fig.~\ref{fig:regionalinertia_39b} show that the effective regional inertia is significantly higher than the conventional regional inertia metric. This indicates that, for this system, the conventional metric systematically underestimates the inertial response. The main reason for this discrepancy is the inter-regional contributions. For example, in Region 5, Fig.~\ref{fig:varyline_region5} illustrates how the effective regional inertia varies as the reactance of the interconnection line 16--19 is scaled by the factor $\alpha_{16-19}$. As the line reactance increases, the proposed regional inertia gradually converges toward the conventional value, emphasizing the role of neighboring inertial responses. In meshed and well-interconnected systems, regions effectively share inertial support, thereby enhancing their effective inertial behavior. This is particularly relevant since some practical methods may misinterpret such enhanced inertia as arising from hidden local resources within the analyzed region.

\subsubsection{Impact of Virtual Inertia Allocation}

To evaluate the impact of virtual inertia allocation on the regional inertial response, a GFM device with virtual inertia capability is connected at bus 4, located in Region 1 (see Fig.~\ref{fig:clusters_39bbench}). We consider the REGFM\_A1 model described in \cite{REGFM_A1}. The P–f droop control scheme is adopted for the GFM device. Under this configuration, the virtual inertia is defined as $H_{GFM} = T_\omega/m_p$, where $T_\omega$ is the time constant of the low-pass filter, and $m_p$ is the P–f droop gain \cite{Eugenie2024}. The internal voltage $E_{GFM}$ of the GFM is regulated via a Q–V droop control. Consequently, to incorporate the GFM into the nodal inertia formulation, the parameters $H_{GFM}$, $E_{GFM}$, and $X_L$ (coupling
reactance) are considered known and can be directly included in equation \eqref{eq:final_h}.

This device can be interpreted as a representative DER equipped with virtual inertia functionality. The GFM unit injects 100 MW of active power into the system. Region 1 is chosen for this study as it contains the highest number of buses and features a well-defined inertial and topological structure that remains unchanged with the addition of the new device. Based on the minimum virtual inertia requirement obtained from equation~\eqref{eq:min_Hj_criterion}, which yields 
$H_{\mathrm{GFM}}^{\mathrm{min}} = 30\,\mathrm{s}$, we define the following three scenarios: (i) $H_{GFM} = 10\,\mathrm{s}$, (ii) $H_{GFM} = 30\,\mathrm{s}$, (iii) and $H_{GFM} = 50\,\mathrm{s}$. The damping added by each case is kept constant ($D=1/m_p=10$).

\begin{figure}[htb]
\includegraphics[width=0.85\linewidth]{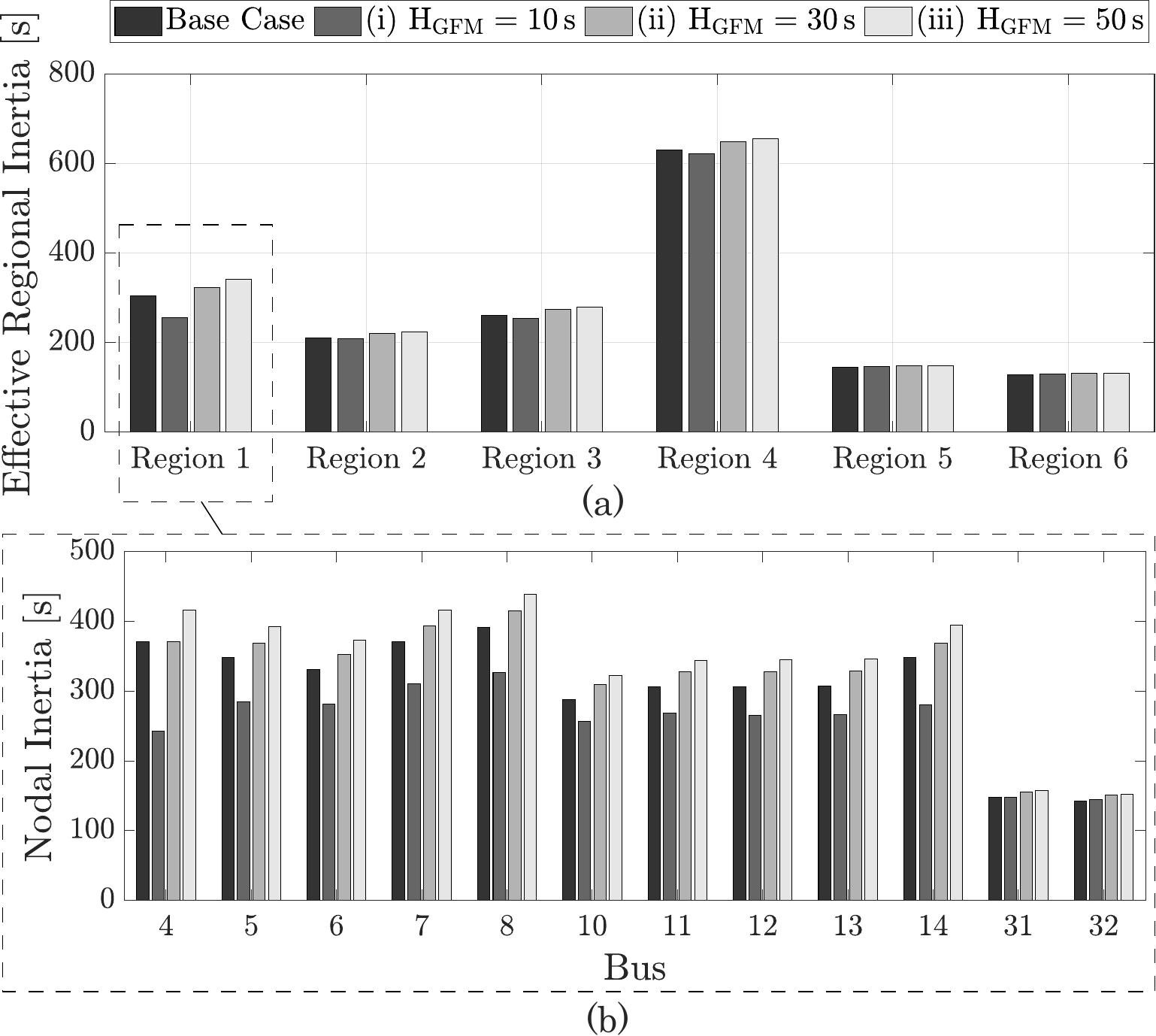}
\centering
\caption{(a) Effective regional inertia across all regions in the system for three virtual inertia scenarios at bus 4, and
(b) Detailed distribution of nodal inertia in Region 01 for the same scenarios.}
\label{fig:regionalinertia_39bGFM}
\end{figure}

The resulting effective regional inertia and the nodal inertia distributions within Region~1 for the three scenarios are shown in Fig.~\ref{fig:regionalinertia_39bGFM}. As expected, scenario~(i) leads to a reduction in both the nodal and effective regional inertia, indicating a degraded inertial response. Scenario~(ii), which meets the minimum requirement, maintains the overall regional inertia close to its original value. Scenario~(iii), with a higher inertia setting, increases both nodal and effective regional inertia, thereby improving the region's inertial support.

It is interesting to note that Region~1 contains two synchronous machines located at Buses~31 and~32, with inertia constants of $30.3\,\mathrm{s}$ and $35.8\,\mathrm{s}$, respectively—values that are close to the minimum requirement obtained from equation~\eqref{eq:min_Hj_criterion}. This observation reinforces the insight presented in Remark~2: the virtual inertia of a new device should ideally be aligned with the inertia characteristics of existing devices within the region. Such alignment helps ensure a balanced and coherent inertial response at the regional level.

\begin{figure}[!htb]
\includegraphics[width=0.85\linewidth]{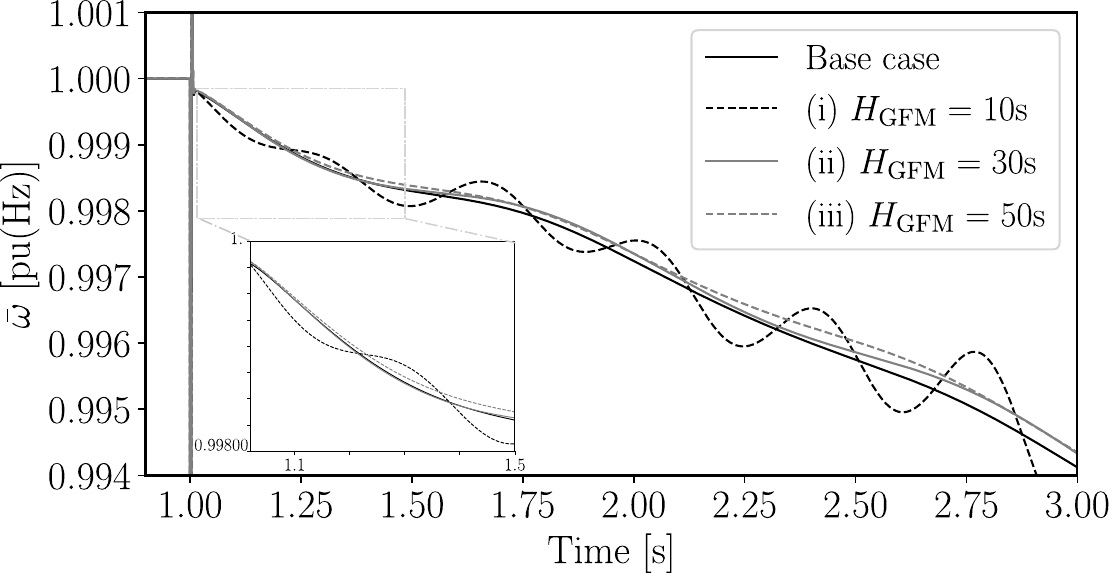}
\centering
\caption{Average regional frequency of Region 1 following a disturbance at bus 10.}
\label{fig:average_freq_region1}
\end{figure}

To assess the dynamic frequency response, the same disturbance used in the last section is applied at bus 10. The frequency at each bus in the region is estimated using a PLL, and the average regional frequency $\bar{\omega}$ is computed for all three scenarios. The results are presented in Fig.~\ref{fig:average_freq_region1}. As anticipated, the use of  $H_{GFM} = 10\,\mathrm{s}$ leads to an increase in the regional RoCoF, indicating a reduction in the effective regional inertia. However, this effect can not be observed if we consider the conventional regional inertia assessment by equation~\eqref{eq:conventional_H}. Furthermore, case (ii) has the same initial frequency response as the base case, showing that the inertial response of the region was not affected by the addition of the new device with $H_{GFM} = H_{\mathrm{GFM}}^{\mathrm{min}} = 30\,\mathrm{s}$.

\vspace{-0.5cm}

\subsection{IEEE 68-Bus System}

The proposed method for quantifying regional inertia is now applied to the IEEE 68-bus system and two modified versions including IBR devices. Generators are modeled using a sixth-order representation with parameters and controls described in \cite{Benchmark}. Three scenarios are considered: (i) base case with 16 synchronous machines; (ii) a GFL inverter without inertial response replacing the synchronous machine at bus 11; and (iii) four GFM inverters, each with $10\,\mathrm{s}$ of virtual inertia, connected at buses 33, 43, 46, and 51.

\begin{figure*}[!htb]
\includegraphics[width=0.9\linewidth]{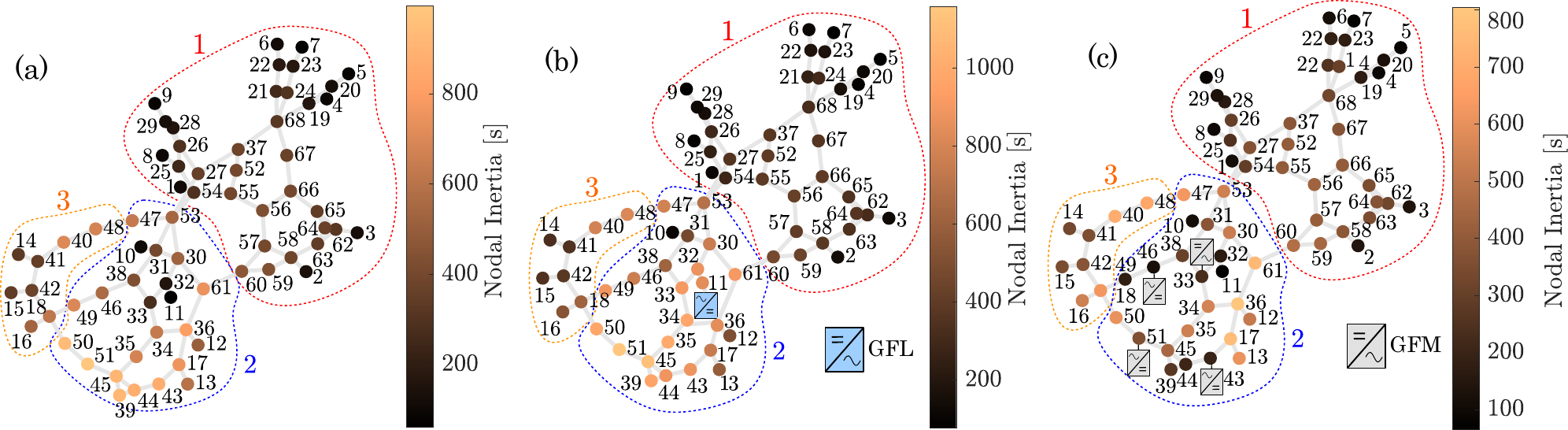}
\centering
\caption{Regional partitions and inertia distribution for: (a) all synchronous machines, (b) synchronous machine at bus 11 replaced by a GFL device, and (c) base case augmented with four GFM devices providing virtual inertia.}
\label{fig:68bus_topo}
\end{figure*}

The network partitioning algorithm identifies three regions. The resulting partitions and inertia distributions are shown in Fig.~\ref{fig:68bus_topo}. Since no topological or significant dispatch changes occur across scenarios, the regional boundaries remain the same.

In the base case, Region 2 exhibits the highest effective regional inertia. Table~\ref{tab:regional_inertia_comparison} compares the proposed effective regional inertia with the conventional approach for this region. The largest increase in effective regional inertia occurs in scenario (ii), where the synchronous machine at bus 11 is replaced by a GFL inverter without inertial response. Interestingly, the conventional method leads to the opposite conclusion, underestimating the replacement’s effect. Moreover, in scenario (iii),  the conventional metric overestimates inertia by assigning the highest $\mathrm{H}_{\mathrm{conv}}$, whereas the proposed $\mathrm{H}_{\mathrm{eff}}$ indicates a reduction.

This result is initially counterintuitive. The synchronous machine at bus 11 has relatively low inertia compared to other generators in its region, introducing a local oscillatory mode that effectively reduces frequency strength. As shown in Fig.\ref{fig:68bus_topo}(a), the nodal inertia at bus 11 is notably small when compared with its region. When replaced by a GFL device, the inertia at the connection bus and neighboring buses becomes dominated by the remaining machines in the region, leading to a significant increase in effective regional inertia (Fig.\ref{fig:68bus_topo}(b)). In scenario (iii), however, the GFM devices are placed at buses that already exhibit high nodal inertia in the base case. In this configuration, maintaining the same inertia level would require substantially higher virtual inertia than the $10\,\mathrm{s}$ assigned (see Remark 2), which is insufficient and results in a decline in the regional inertia distribution.

\begin{table}[htb!]
\centering
\caption{Regional Inertia of Area 2 for Different Scenarios: Effective and Conventional Methods -- IEEE 68-bus.}
\begin{tabular}{ccccc}
\hline
\hline
\multirow{2}{*}{Scenario} & $\mathrm{H}_{\text{eff}}$ (s) & $\Delta \mathrm{H}_{\text{eff}}$ (s) & $\mathrm{H}_{\text{conv}}$ (s) & $\Delta \mathrm{H}_{\text{conv}}$ (s) \\ 
 & & (from base) & & (from base) \\ \hline
(i) & 601.7 & -- & 399.5 & -- \\
(ii) & 793.9 & +192.2 & 371.3 & -28.2 \\
(iii) & 413.8 & -187.9 & 439,5 & +40 \\ \hline\hline
\end{tabular}
\label{tab:regional_inertia_comparison}
\end{table}

\begin{figure}[!htb]
\includegraphics[width=0.8\linewidth]{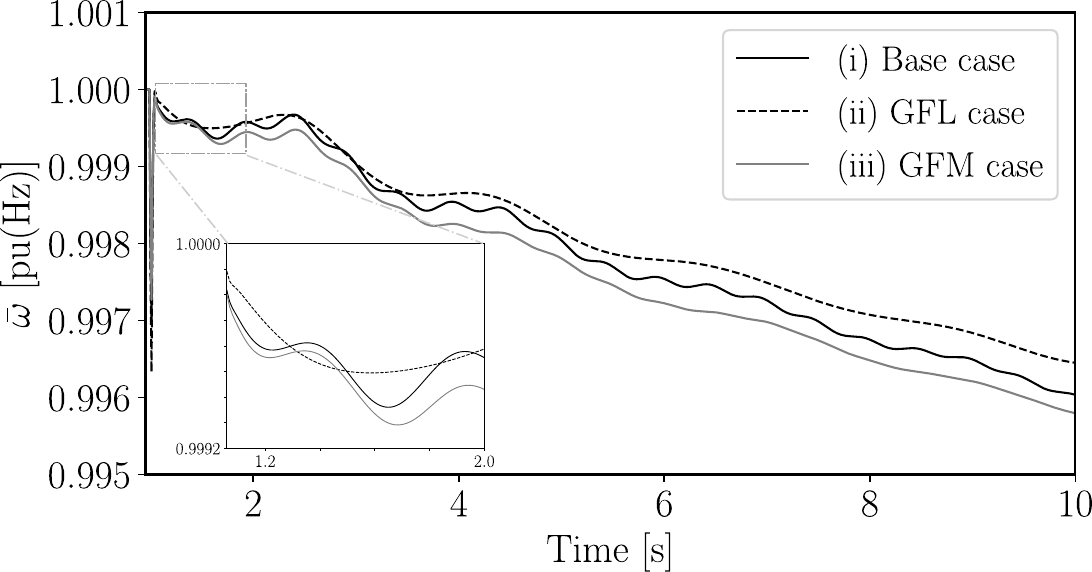}
\centering
\caption{Average regional frequency of Region 2 following a disturbance at bus 30.}
\label{fig:average_freq_region2_68b}
\end{figure}

Dynamic validation is performed by applying a 5~pu load step at bus 30. The average frequency of Region 2 is shown in Fig.\ref{fig:average_freq_region2_68b}. Scenario (ii) exhibits the lowest frequency variability after the disturbance, consistent with the increase in effective regional inertia. In contrast, scenarios (i) and (iii) present local oscillatory modes that degrade both RoCoF and frequency variability, while these oscillations are absent in scenario (ii).

These findings emphasize that regional inertia cannot be reliably assessed using conventional aggregation metrics. In particular, a low-inertia synchronous machine may introduce unfavorable local oscillatory behavior that reduces the effective regional inertia, even though the conventional method suggests otherwise. The analysis further shows that GFL inverters, especially when combined with fast frequency response (FFR), are preferable in two situations: (i) when replacing low-inertia machines in regions with relatively high effective regional inertia; and (ii) when connected to buses with relatively high nodal inertia levels.

\vspace{-0.3cm}

\section{Conclusion}

This work establishes a novel framework for assessing effective regional inertia in power systems. Leveraging the spatial distribution of inertia and system connectivity, a network partitioning methodology was developed to define coherent areas. Within each region, an effective regional inertia metric was introduced based on the average nodal inertia. Unlike traditional methods that aggregate regions into equivalent generators, the proposed formulation preserves local dynamics and provides a more accurate representation of regional frequency behavior. Simulation results on the IEEE 39-bus and IEEE 68-bus systems demonstrate that the integration of inertial devices may not necessarily enhance the regional inertia. The allocation of new inertial devices should therefore consider both the regional inertia characteristics and the system topology, which together determine the extent to which the added inertia contributes to improve the regional frequency dynamics. Future research will focus other applications using the effective regional inertia quantification, e.g., equivalents of large power systems.

\vspace{-0.3cm}

\bibliographystyle{bibliography/IEEEtran}

\end{document}